\documentclass[12pt]{article}
\begin{document}
\newcommand{\br}{{\bf r}}
\newcommand{\grad}{\mbox{\boldmath$\nabla$}}
\newcommand{\bdiv}{\mbox{\boldmath$\nabla\cdot$}}
\newcommand{\curl}{\mbox{\boldmath$\nabla\times$}}
\newcommand{\bcdot}{\mbox{\boldmath$\cdot$}}
\newcommand{\btimes}{\mbox{\boldmath$\times$}}
\newcommand{\btau}{\mbox{\boldmath$\tau$}}
\newcommand{\btheta}{\mbox{\boldmath$\theta$}}
\newcommand{\bmu}{\mbox{\boldmath$\mu$}}
\newcommand{\bepsilon}{\mbox{\boldmath$\epsilon$}}
\newcommand{\bcj}{\mbox{\boldmath$\cal J$}}
\newcommand{\bcf}{\mbox{\boldmath$\cal F$}}
\newcommand{\bbeta}{\mbox{\boldmath$\beta$}}
\newcommand{\lbar}{\lambda\hspace{-.09in}^-}
\newcommand{\bcp}{\mbox{\boldmath$\cal P$}}
\newcommand{\bco}{\mbox{\boldmath$\omega$}}
\newcommand{\brho}{\mbox{\boldmath$\rho$}}
\newcommand{\balpha}{\mbox{\boldmath$\alpha$}}
\title{Infinite self energy of a point charge?}
\author{Jerrold Franklin\footnote{Internet address:
Jerry.F@TEMPLE.EDU}\\
Department of Physics\\
Temple University, Philadelphia, PA 19122-6082}
\date{\today}
\maketitle
\begin{abstract}
The question of whether a point charge has infinite self energy is investigated.
We find that a point charge does not have any electromagnetic self energy.
\end{abstract}
\section{Introduction}

The notion of an infinite electromagnetic self energy of point charges (presumably electrons) is often assumed. See, for instance,\cite{dg,rf,jdj}. However, each of these sources acknowledge that they don't understand that result.
For instance, Giffiths writes, ``The infinite energy of a point charge is a recurring source of embarrassment
for electromagnetic theory".\footnote{\cite{dg}, page 96.},
and Feynman states,\footnote{\cite{rf}, 8-6.} 
``We must conclude that the idea of locating the energy in the field is inconsistent with the assumption of the existence of point charges."
 In this paper, we show that electrons must be point particles with no electromagnetic self energy.  

\section{Energy of discrete electrons} 

To study charge distributions in classical electromagnetism, we consider only electrons because there are no known fundamental particles whose charge's magnitude is different than that of an electron or an integral multipole of its charge. (If free quarks were found to exist, they could be included.)

For many years there have been attempts, notably by Lorentz\cite{hl} 130 years ago, to consider electrons as charged particles with finite or infinitesimal radius. This failed because such an electron is known to have a measured radius of 
at most $10^{-22}m.$\footnote{ Observation of a single electron in a Penning trap\cite{pt} suggests the upper limit of the particle's radius to be 10{-22} m.}
If it were considered a uniformly charged sphere of that radius, 
it would have a self energy of\footnote{We are using Gaussian units with ${\hbar}$ and $c=1$.}  
$W=\frac{e^2}{R}=2\times 10^{24}$ MeV.

A collection of N individual electrons has Coulomb energy\footnote{This equation is equivalent to Eq,~(1.51) of\cite{jdj}.}
\begin{equation}
W=\frac{e^2}{2}\sum_i^N\sum_{j\ne i}^N\frac{1}{|{\bf r}_i-{\bf r}_j|}. 
\end{equation}
The condition, $j\ne i$ is necessary because the first electron put in place has no Coulomb energy. This equation demonstrates that none of the electrons in a discrete sum has a one particle energy that could be considered self energy.  That should be enough, but we have to go on.

\section{Electromagnetic energy of a `continuous' electron charge distribution.}
Because it is clear from Eq.~(1) that any electron in a discrete sum of electrons has no electromagnetic self energy, many sources have tried to include the self energy in a `continuous' charge distribution of electrons. 
The reason we have put continuous in quotes is that there is, in fact, no possible continuous charge distribution of electrons. 

The standard way to convert the discrete sum in Eq.~(1) into a continuous distribution is to take the limits,\footnote{In reference\cite{dg}, Griffiths also has made the point that the charge at a particular point must be infinitesimal to achieve a continuous charge distribution.}
\begin{equation}
 N\rightarrow\infty,\quad e\rightarrow 0.
\end{equation}
However, electrons have a constant charge, $e$, that cannot be reduced or considered infinitesimal. 

 That means that there can be no truly continuous charge distribution in classical electromagnetism. Actually, that is not a serious problem because leaving $N$ finite, but very large (of the order of Avogradro's constant), seems to work. If N is that large, calling the sum in Eq.~(1) `continuous' does no harm, {\it unless} it is used to invoke electromagnetic `self energy'.

We also see that the $\rho({\bf r})$ and ${\bf j(r)}$\footnote{We have shown in \cite{j} that an electric current density is actually composed of moving electrons, and is not strictly continuous.} 
  in Maxwell's equations, are not exactly continuous. That does no harm ({\it unless} it is used to invoke electromagnetic `self energy'.) if N is large enough.
 Some sources have used their particular mistakes to produce elactromagnetic self energy for a point charge in a `continuous' distribution. We discuss some of these mistakes in the next section.
 
\section{Producing `Infinite Self Energy'}

Jackson\cite{jdj} takes Eq.~(1) to the continuom limit (as described above) to give the equation,\footnote{This is Eq.~(4.83) of \cite{jdj}.} 
\begin{equation}
 W=\frac{1}{2}\int\rho({\bf r})\phi({\bf r})d^3r
 =-\frac{1}{8\pi}\int\phi({\bf r})\nabla^2\phi({\bf r})d^3r.
\end{equation}
Then, he goes on to say, ``Integration by parts leads to the result,
 \begin{equation}
 W=\frac{1}{8\pi}\int|\grad\phi({\bf r}|^2 d^3r
 =\frac{1}{8\pi}\int|{\bf E( r)}|^2 d^3r,
\end{equation}
where the integral is over all space.'', He is considering charges at rest, so there is no kinetic energy, and the electric field is given
by $\bf E=-\grad\phi$, with no contribution from the vector potential, $\bf A$.

There is ordinarily a surface integral on the right hand side of Eq.~(4), because integration by parts gives
 \begin{equation}
 W=-\frac{1}{8\pi}\int\phi({\bf r})\nabla^2\phi({\bf r})d^3r.
=\frac{1}{8\pi}\int|\grad\phi({\bf r}|^2 d^3r,
  -\oint_S\bf dS\bcdot(\grad\phi)\phi.
\end{equation}
The surface integral vanishes for an infinite volume because $\bf E$ is assumed to vanish like $\frac{1}{r^2}$ as $r\rightarrow\infty$. 

Then, Jackson claims that, ``This leads naturally to the identification of the integrand as an energy density, $w$:
\begin{equation}
 w=\frac{1}{8\pi}|{\bf E}|^2.
\end{equation}
But, Eq.~(4) {\it does not} lead to an energy density because an integral over all space does not define an energy density. An energy density is a function whose intregal over ANY 
volume, not just ``all space'', gives the energy within that  volume. For a finite volume, the surface integral in Eq,~(5) must be added to Eq,~(4).

Jackson's whole derivation, and its conclusion has to be redone correctly. That is especially true for a point charge if you want to discuss its self energy.
The first important point is that, if $\rho$ is a point charge, the potential should be written as $\phi_{\rm other}$ to show that the potential does not include the point charge acting on itself.
That corresponds to the subscript, $j\ne i$ in Eq.~(1).

This means that Eq.~(3) for a point charge, $q$, at a position, $\bf r_0$, should be written,
\begin{eqnarray}
 W_q&=&\frac{1}{2}q\phi_{\rm other}({\bf r_0})\nonumber\\
&=&\frac{q}{2}\int_V\delta({\bf r-r_0})\phi_{\rm other}({\bf r}))d^3r,
\quad{\rm where, \,V\,must\,include\,the\,position\,{\bf r_0}}.
\nonumber\\
&=&-\frac{q}{8\pi}\int\phi_{\rm other}({\bf r})\nabla^2\left(\frac{1}{|{\bf r-r_0}|}\right)d^3r
\nonumber\\
&=&-\frac{1}{8\pi}\int\left\{\grad\bcdot\left[\phi_{\rm other}({\bf r})\grad\left(\frac{q}{|{\bf r-r_0}|}\right)
-\left[\grad\phi_{\rm other}({\bf r})\right]\bcdot\grad\left(\frac{q}{|{\bf r-r_0}|}\right)\right]\right\}d^3r\nonumber\\
&=&\frac{1}{8\pi}\int\left[\grad\phi_{\rm other}({\bf r})\right]\bcdot
\left[\grad\left(\frac{q}{|{\bf r-r_0}|}\right)\right]d^3r
-\oint\bf dS\bcdot
\left[\grad\left(\frac{q}{|{\bf r-r_0}|}\right)\right]\phi_{\rm other}({\bf r})\nonumber\\
&=&\frac{1}{8\pi}\int_V({\bf E}_{\rm other}\bcdot{\bf E}_q)d^3r
+\oint_S\bf dS\bcdot{\bf E}_q\phi_{\rm other}.
\end{eqnarray}
It is clear from the above derivation that ${\bf E}_q$ and  ${\bf E}_{\rm other}$ are different fields.
The point charge is only in ${\bf E}_q$, and not 
in ${\bf E}_{\rm other}$, so there can be no self energy in Jackson's model.

@flatwarE137
Feynman starts his derivation of the self energy  of a point charge by starting with the same equation as Jackson, 
\begin{equation}
 W=\frac{1}{2}\int\rho({\bf r})\phi({\bf r})d^3r.
\end{equation}
Then he makes the same mistake by not realizing that, for a point charge, it should be written as
\begin{eqnarray}
 W_q&=&\frac{1}{2}q\phi_{\rm other}({\bf r_0}).
 \end{eqnarray}
 Because he doesn't recognize this, he gives the same integral as Jackson,\footnote{This is equation (8.36) of \cite{rf}.}
 \begin{equation}
 W=\int{\bf E\bcdot E}d^3r
 =\int_0^\infty \frac{q^2dr}{2r^2}=\frac{q^2}{2r}{\bigg|}_0^\infty.
 \end{equation}

 He then spends one paragraph trying to understand this infinite result, coming to the conclusion: ``We must conclude that the idea of locating the energy in the field is inconsistent with the assumption of the existence of point charges.''

\section{Conclusion}

We have shown that there is no infinite self energy for a point charge in a discrete collection of point charges, or what has been called a `continuous' charge distribution in classical electromagnetism. In Section 4, we have shown that two leading attempts to produce infinite self energy of a point charge are wrong.

\end{document}